\documentclass[aps,preprint,floatfix,nofootinbib,showpacs]{revtex4-1}
\pdfoutput=1
\usepackage{graphicx,color}
\usepackage{hyperref}
\usepackage{amsmath}
\usepackage{amsfonts}
\usepackage{amssymb}
\usepackage{setspace}

\begin{document}
\title{Interpreting the 750 GeV Di-photon Resonance using photon-jets
in Hidden-Valley-like models}

\renewcommand{\thefootnote}{\arabic{footnote}}

\author{
Jung Chang$^1$, Kingman Cheung$^{1,2,3}$, and Chih-Ting Lu$^3$
}
\affiliation{
$^1$ Physics Division, National Center for Theoretical Sciences,
Hsinchu, Taiwan \\
$^2$ Division of Quantum Phases and Devices, School of Physics, 
Konkuk University, Seoul 143-701, Republic of Korea \\
$^3$ Department of Physics, National Tsing Hua University,
Hsinchu 300, Taiwan
}
\date{\today}

\begin{abstract}
Motivated by the di-photon resonance recently reported by the ATLAS and
CMS collaborations at $ \sqrt{s} =13$ TeV, we interpret the resonance
as a scalar boson $X (750)$ in hidden-valley-like models.
The scalar boson $X$ can mix with the standard model  Higgs boson and thus can
be produced via gluon fusion. It then decays into a pair of very light
hidden particles $Y$ of $\mathcal{O}(1\,{\rm GeV})$, 
each of which in turn decays to a 
pair of collimated $ \pi^{0} $'s, and these two $ \pi^{0} $'s
decay into photons which then form photon-jets.
A photon-jet ($ \gamma $-jet) is a special feature that consists 
of a cluster of collinear photons from the decay of a fast moving 
light particle ($\mathcal{O}(1\,{\rm GeV})$). 
Because these photons inside the photon-jet are so
collimated that it cannot be distinguished from a single photon,
and so in the final state of the decay of $X(750)$ a pair of photon-jets
look like a pair of single photons, 
which the experimentalists observed and formed the 750 GeV di-photon 
resonance.
Prospects for the LHC Run-2 about other new and testable 
features are also discussed.
\end{abstract}

\maketitle

\section{Introduction}

The Run-2 of the LHC has caught the eyes of everyone with a potential
particle at around 750 GeV.
Both the ATLAS \cite{ATLAS-750} and CMS \cite{CMS-750}
collaborations have reported a "bump" in the
di-photon invariant mass distribution around 750 GeV.
The ATLAS Collaboration with a luminosity of 3.2 fb$^{-1}$ showed a resonance
structure at $M_X \approx 750$ GeV with a local significance of $\sim 3.64
\sigma$ but corresponding to $1.88\sigma$ when the look-elsewhere-effect is
taken into account \cite{ATLAS-750}. 
The CMS Collaboration also reported a similar though smaller 
excess with a luminosity of 2.6 fb$^{-1}$ at $M_X \approx 760 $ GeV with
a local significance of $2.6\sigma$ but a global significance less than
$1.2\sigma$ \cite{CMS-750}.  
Also, in the analysis of ATLAS a total width of
$ \Gamma/M\approx 6\% \sim 45$ GeV is preferred \cite{ATLAS-750}.
Such a large width might  indicate something new beyond
the Standard Model.
Let us summarize the data here:
\begin{eqnarray}
{\rm ATLAS} &:& M_X = 750 \;{\rm GeV},\;\;\; 
 \sigma_{\rm fit} (pp \to X \to\gamma\gamma) \approx 10 \pm 3\,
{\rm fb}\;(95\% \;{\rm CL}),\;\; \Gamma_X \approx 45 \,{\rm GeV} \nonumber \\
{\rm CMS} &:& M_X = 760 \,{\rm GeV} ,\;\;\;
 \sigma_{\rm fit} (pp \to X \to\gamma\gamma) \approx 9 \pm 7\,
{\rm fb}\;(95\% \;{\rm CL}) \;.
\end{eqnarray}
The uncertainties shown are $1.96\sigma$ corresponding to 95\% CL.
Note that we estimate the best-fit cross section from the 95\%CL upper
limits given in the experimental paper, by subtracting the ``expected'' 
limit from the ``observed'' limit at $M_X = 750\; (760)$ GeV for ATLAS
(CMS). 

Before further exploring, let us summarize the information that is
available so far to figure out some characteristics of new resonance.
First, because the bump was observed in the di-photon
channel, it forbade the direct decay of an on-shell $spin-1$ particle into
di-photons by the Landau-Yang theorem
\cite{Landau:1948kw}\cite{Yang:1950rg}.
\footnote{
  It is nevertheless possible for a spin-1 particle to decay into
two photon-jets via cascade decays, such that the spin-1 resonance for
the 750 GeV resonance is not entirely excluded.
}
The spin of this new
particle turns out to be either $spin-0$ or $spin\geq 2$. 
Second, because the production cross section is rather large 
$ \mathcal{O}(10\, {\rm fb}) $, 
and the parton luminosity of gluons increases faster than that of quarks
\cite{PDF}, the production mode is more
likely the gluon fusion rather than quark-antiquark annihilation. 
Finally, because the total width of the new resonance is rather
large ($ \Gamma/M\approx 6\% $), it implies that 
the new particle couples rather strongly to its decay products or
decays into objects that are hard to be observed.
Given the severe exclusion bounds in search of new resonances decaying
into vector-boson pairs, di-leptons, dijets, and $t\bar t$ pairs
\cite{PDG,ATLAS-CMS constraint}, the new particle at 750 GeV 
may strongly couple to its decay particles, which 
are hard to be observe or excluded and also the decays into 
SM particles must be suppressed.

Although the hint is very preliminary, it has stimulated a lot of
phenomenological activities, bringing in a number of models for 
interpretation.  The first category is the Higgs section extensions, 
including adding singlet Higgs fields \cite{singlet,hidden,dm}, 
two-Higgs-doublet models
and the MSSM \cite{2hdm}, but in general it fails to explain the large 
production cross section of $pp \to H \to \gamma \gamma$ in the conventional 
settings, unless additional particles are added, for example, 
vector-like fermions \cite{2hdm,singlet,hidden,dm}. Another
category is the composite models \cite{composite} that naturally contain heavy 
fermions, through which the production and the diphoton decay of the 
scalar boson can be enhanced.  Other possibilities are also entertained, such
as axion \cite{axion}, sgoldstini \cite{goldstino}, radion \cite{radion}. 
More general discussion can be found in Refs.~\cite{general}.
The generic feature of the suggested interpretations is to enhance
the production cross section of $pp \to H \to \gamma\gamma$, where $H$
is the 750 GeV scalar or pseudoscalar boson, by additional particles running
in the $H\gamma\gamma$ decay vertex and/or $Hgg$ production vertex.
Another generic feature though not realized in the CMS data is the relatively
broad width of the particle, which motivates the idea that this particle
is a window to the dark sector or dark matter \cite{dm,hidden}.

For simplicity we consider the resonance to be a $spin-0$ particle and
produced by gluon fusion. To make this $spin-0$ particle strongly
couple to its decay products (which are hard to be observed) 
and suppress its decays into $WW$, $ZZ$, $\ell^+\ell^-$, $t\bar t$ and
dijet pair but still can accommodate the experimental results, 
we advocate a special scenario: 
{\it
 Could the final state be photon-jets instead of a pair of single photons?
} 
 
A photon-jet ($ \gamma $-jet) \cite{photon-jet} is a special feature
that consists of a cluster of collinear photons from the decay of a
fast moving light particle ($\mathcal{O}(1\,{\rm GeV})$). 
Because these photons inside the
photon-jet are so collimated that they cannot be distinguished from a single
photon, so a photon-jet looks like a photon experimentally unless specific
procedures to unlock the substructure inside the photon-jet.
We propose a {\it Hidden-Valley-like simplified model} 
\cite{Hidden Valley model}, in which the $spin-0$ particle $ X(750) $ 
decays to a pair of very light ($\mathcal{O}(1\,{\rm GeV})$) 
particles $ Y $ 
(which strongly couples to $ X $). Each $ Y $ in turn decays to a pair of 
collimated neutral pions $ \pi^{0}$, and these two $ \pi^{0} $ decay to 
photons. Therefore, each fast moving $Y$ gives rise to a photon-jet, and 
in the final state of the decay of $X(750)$ consists of a pair of photon-jets,
which the experimentalists observed and formed the 750 GeV di-photon 
resonance.

The organization is as follow. We describe a Hidden-Valley-like
simplified model in the next section, and in Sec. III the existing
constraints for $X(750)$. Then we fit the di-photon resonance in the
model in Sec. IV. Finally, we provide some discussion and outlook in Sec. V.

\section{Hidden-Valley-like Simplified Model}

\begin{figure}[t!]
\centering
\includegraphics[width=3.5in]{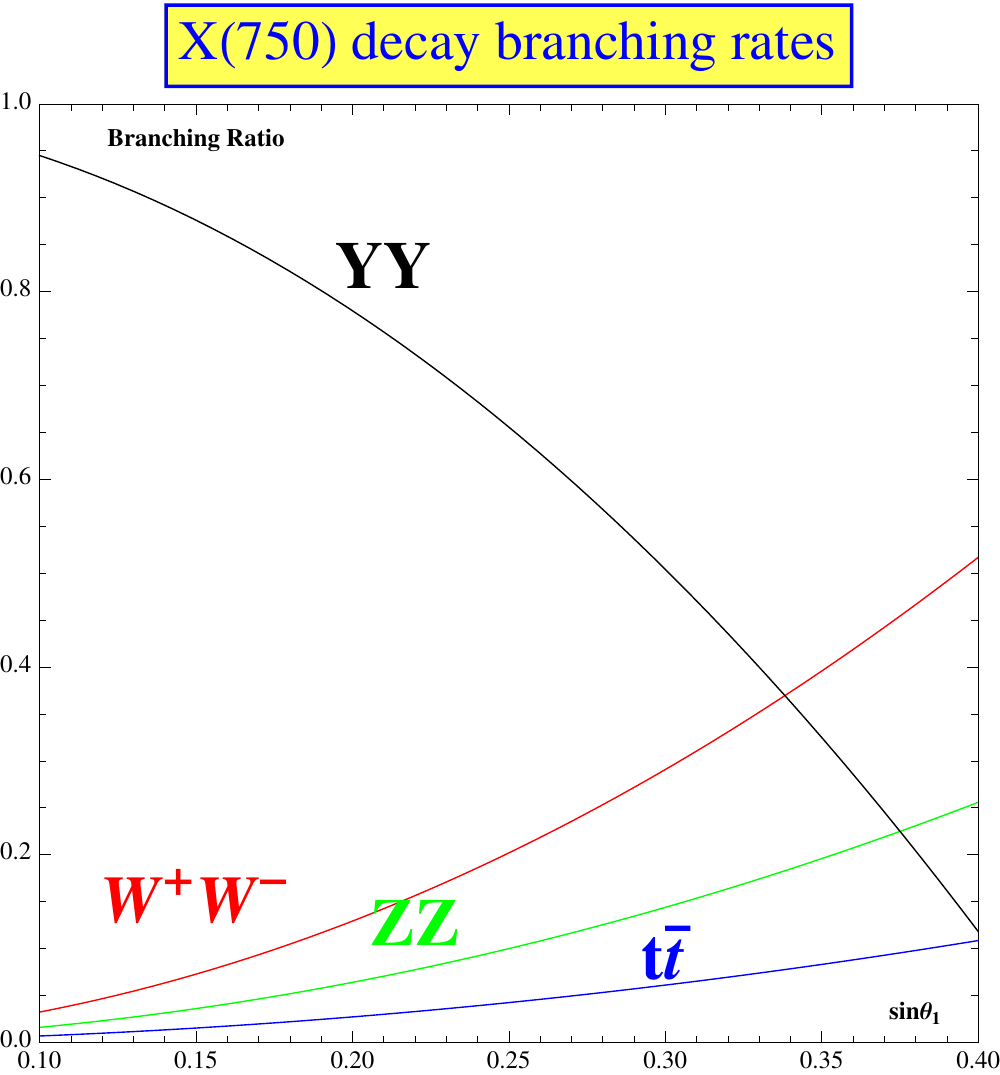}
\caption{\small \label{X-BR}
The branching ratios of the scalar boson $ X(750) $ with 
$ \Gamma = 45 GeV $ for the four most dominant modes 
$ YY $, $ W^{+}W^{-} $, $ ZZ $, and $ t\overline{t} $ }
\end{figure}

Here we employ a Hidden Valley-like model \cite{Hidden Valley model}
in which the SM Higgs field $\Phi$ can mix with two real scalar fields
$\chi_{1}$ and $\chi_{2}$. These additional scalar fields 
are neutral under the SM gauge group $ G_{SM} $ and do not have
any SM interactions. The simplified Lagrangian for this model is given
by
\footnote{
Since we propose a simplified model, we do not take into account
the details of the vevs for these singlet fields, their vevs will be
related to the parameters $M, N$ in our simplified model.}
\begin{eqnarray}
{\cal L} &=& \frac{1}{2}\partial_{\mu}\chi_{1}\partial^{\mu}\chi_{1}
 +\frac{1}{2}\partial_{\mu}\chi_{2}\partial^{\mu}\chi_{2}
+\frac{1}{2}\mu^{2}_{1}\chi^{2}_{1}+\frac{1}{2}\mu^{2}_{2}\chi^{2}_{1}
 +\mu^{2}_{3}\chi_{1}\chi_{2} \nonumber \\
&+& (\lambda_{1}\chi_{1}+\lambda_{2}\chi_{2})[M(\Phi^{\dagger}\Phi)
  +N(\lambda_{1}\chi_{1}+
\lambda_{2}\chi_{2})^{2}] \nonumber \\
&+& {\cal L_{SM}} \;, 
\label{L'}
\end{eqnarray}
where the Higgs sector in the $ {\cal L_{SM}} $ is
\begin{eqnarray}
{\cal L_{SM}} &\supset& (D_{\mu}\Phi)^{\dagger}(D^{\mu}\Phi)
 +\mu^{2}(\Phi^{\dagger}\Phi)-\lambda (\Phi^{\dagger}\Phi)^{2} \;.
\end{eqnarray}
In the Lagrangian,
all $\mu_i$, $ M $ and $ N $ are of mass dimension 1 
while $\lambda$s are of mass dimension 0.
After the electroweak symmetry breaking (EWSB), the SM Higgs 
doublet field $\Phi$ is expanded around its vacuum-expectation value:
\begin{equation}
\Phi = \frac{1}{\sqrt{2}} \left( \begin{array}{c}
           0 \\
           \langle \phi \rangle + \phi(x) \end{array} \right ) \;,
\end{equation}
where $\langle\phi\rangle\approx 246$ GeV.
It is easy to see that the Higgs boson $\phi$ will mix with these two
new scalar bosons $\chi_{1}$ and $\chi_{2}$ to form mass eigenstates 
denoted by $h$, $X$ and $Y$,
respectively. The mass terms for the Higgs boson and these two new scalar 
bosons are
\begin{equation}
{\cal L}_m =  - \frac{1}{2} \left( \phi \; \chi _{1} \; \chi _{2} \right )\,
 \left( \begin{array}{ccc} 
             2\lambda\langle\phi\rangle^{2}  
& -\lambda_{1}M\langle\phi\rangle & -\lambda_{2}M\langle\phi\rangle \\
            -\lambda_{1}M\langle\phi\rangle & -\mu_1^2 & -\mu_3^2 \\
            -\lambda_{2}M\langle\phi\rangle & -\mu_3^2 & -\mu_2^2           
                                       \end{array} \right )\,
  \left( \begin{array}{c} 
           \phi \\
           \chi _{1} \\
           \chi _{2} \end{array} \right ) \;,
\end{equation}
We can rotate $(\phi\; \chi_{1}\; \chi_{2}) \longrightarrow (h \; X \; Y)$ 
through these angles $ \theta_{1} $, $ \theta_{2} $ and $ \theta_{3} $
\begin{equation}
 \left( \begin{array}{c}
                h \\ 
                X \\
                Y \end{array} \right ) 
= 
 \left( \begin{array}{ccc}
           \cos\theta _{1} & \sin\theta _{1} & 0 \\
          -\sin\theta _{1} & \cos\theta _{1} & 0 \\
           0 & 0 & 1 \end{array}
 \right )\,
 \left( \begin{array}{ccc}
           \cos\theta _{2} & 0 & \sin\theta _{2} \\
           0 & 1 & 0 \\
          -\sin\theta _{2} & 0 & \cos\theta _{2} \end{array}
 \right )\,
 \left( \begin{array}{ccc}
           1 & 0 & 0 \\
           0 & \cos\theta _{3} & \sin\theta _{3} \\
           0 & -\sin\theta _{3} & \cos\theta _{3} \end{array}
 \right )\, 
 \left( \begin{array}{c}
                \phi \\ 
                \chi _{1} \\
                \chi _{2} \end{array} \right ) 
\end{equation}
\begin{equation}
 = 
 \left( \begin{array}{ccc}
           C_{\theta _{1}}C_{\theta _{2}} & (S_{\theta _{1}}C_{\theta _{3}}-C_{\theta _{1}}S_{\theta _{2}}S_{\theta _{3}}) & (S_{\theta _{1}}S_{\theta _{3}}+C_{\theta _{1}}S_{\theta _{2}}C_{\theta _{3}}) \\
           -S_{\theta _{1}}C_{\theta _{2}} & (C_{\theta _{1}}C_{\theta _{3}}+S_{\theta _{1}}S_{\theta _{2}}S_{\theta _{3}}) & (C_{\theta _{1}}S_{\theta _{3}}-S_{\theta _{1}}S_{\theta _{2}}C_{\theta _{3}}) \\
           -S_{\theta _{2}} & -C_{\theta _{2}}S_{\theta _{3}} & C_{\theta _{2}}C_{\theta _{3}} \end{array}
 \right )\,
 \left( \begin{array}{c}
                \phi \\ 
                \chi _{1} \\
                \chi _{2} \end{array} \right ) 
\end{equation}
where 
$ \theta_{1,2,3} $ is the mixing angle between $ \phi $ and $ \chi _{1} $, 
between $ \phi $ and $ \chi _{2} $, and between 
$ \chi _{1} $ and $ \chi _{2} $, respectively.
$ C_{\theta_{i}} $ stands for $\cos \theta_{i} $ and 
$ S_{\theta_{i}} $ stands for $\sin \theta_{i} $.
If we assume $\theta_{1}$, $\theta_{2}$ are rather small compared with 
$\theta_{3}$, then
the masses for the Higgs boson $h$ and two scalar bosons $X$, $Y$,
and the interaction governing $X \to h h$, $X \to Y Y$
are given by, in terms of the parameters in Eq.~(\ref{L'}):
\begin{eqnarray}
m_h^2 & \simeq &  2\lambda\langle\phi\rangle^{2} -
(\mu^{2}_{1}sin^{2}\theta_{1}+
\mu^{2}_{2}sin^{2}\theta_{2})-M\langle\phi\rangle(\lambda_{1}sin2\theta_{1}+
\lambda_{2}sin2\theta_{2}) \nonumber\\  
&&= (125\;{\rm GeV} )^2 \nonumber \\
m_{X}^2 & \simeq & -\mu^{2}_{1}cos^{2}\theta_{3}-
\mu^{2}_{2}sin^{2}\theta_{3}-
\mu^{2}_{3}sin2\theta_{3} 
+2\lambda\langle\phi\rangle^{2}sin^{2}\theta_{1} +
\lambda_{1}M\langle\phi\rangle \sin2\theta_{1}
\nonumber\\
m_{Y}^2 & \simeq & 
-\mu^{2}_{1}sin^{2}\theta_{3}-
\mu^{2}_{2}cos^{2}\theta_{3}+
\mu^{2}_{3}sin2\theta_{3} 
+2\lambda\langle\phi\rangle^{2}sin^{2}\theta_{2} +
\lambda_{2}M\langle\phi\rangle \sin2\theta_{2}
\nonumber\\
{\cal L}_{Xhh} &\simeq & \frac{1}{2}[  2\lambda\langle\phi\rangle cos^{2}\theta_{1} \sin\theta_{1} +
6N\lambda^{3}_{1}cos\theta_{1}sin^{2}\theta_{1} 
\nonumber\\
&&+\lambda_{1}M(\cos^{3}\theta_{1}-2\cos\theta_{1} \sin^{2}\theta_{1})] Xhh\equiv
\frac{\mu_{Xhh}}{2}Xhh
\label{LXhh} \\
{\cal L}_{XYY} &\simeq & \frac{6N}{2}[\lambda^{3}_{2}\cos ^{2}\theta_{3}\sin\theta_{3}+\lambda^{3}_{1}
\cos\theta_{3}\sin^{2}\theta_{3}+\lambda_{1}
\lambda_{2}^{2}(\cos^{3}\theta_{3}-2\cos\theta_{3} \sin^{2}\theta_{3})\nonumber\\ 
&&+\lambda_{2}\lambda_{1}^{2}(\sin ^{3}\theta_{3}-2\cos^{2}\theta_{3} \sin\theta_{3})]XYY\equiv\frac{\mu_{HS}}{2}XYY
\label{LHD} \;,
\end{eqnarray}
In order to interpret the 750 GeV di-photon resonance, we set $m_{X} = 750$ GeV.
The 750 GeV scalar boson $ X $ can decay into SM particles
via the mixing with the SM Higgs boson. Thus, the partial decay widths for 
$X\rightarrow W^{+}W^{-} $, $ X\rightarrow ZZ $ and $ X\rightarrow
t\overline{t} $ are given by \cite{Gunion:1989we}
\begin{eqnarray}
\Gamma ( X \to W^{+}W^{-}) &=& \sin^2\theta_{1} \, 
  \frac{g^{2}}{64\pi}\frac{m_{X}^{3}}{m_{W}^{2}}\sqrt{1-\frac{4 m_W^2}{m_X^2}}
\left( 1 - 
  \frac{4 m_W^2}{m_X^2}+\frac{12 m_W^4}{m_X^4}\right)\;, \\
\Gamma ( X \to ZZ) &=& \sin^2\theta_{1} \, 
  \frac{g^{2}}{128\pi}\frac{m_{X}^{3}}{m_{Z}^{2}}\sqrt{1-\frac{4 m_Z^2}{m_X^2}}
\left( 1 - 
  \frac{4 m_Z^2}{m_X^2}+\frac{12 m_Z^4}{m_X^4}\right)\;, \\
\Gamma ( X \to t\overline{t}) &=& \sin^2\theta_{1} \, 
  \frac{N_{c}g^{2}m_{t}^{2}}{32\pi m_{W}^2}\left( 1 - 
  \frac{4 m_t^2}{m_X^2}\right)^{3/2}(1+\triangle_{QCD})\;, 
\end{eqnarray}    
where $ \triangle_{QCD} $ are higher order QCD corrections 
\cite{Djouadi:2005gi}.
Other than the decays into SM particles via the mixing with the Higgs 
boson, $X(750)$ can have more decay channels 
$ X\rightarrow YY $ and $ X\rightarrow hh $. 
The partial decay width for $ X\rightarrow YY $ is given by
\begin{eqnarray}
\Gamma (X \to YY) &=& \frac{\mu _{HS}^{2}}{32\pi m_{X}}\times 
    \sqrt{1-4 \left( \frac{m_{Y}}{m_{X}} \right)^{2}} \;.
\end{eqnarray}
and that for $ X\rightarrow hh $ is given by
\begin{eqnarray}
\Gamma (X \to hh) &=& \frac{\mu_{Xhh}^{2}}{32\pi m_{X}}\times 
\sqrt{1-4 \left( \frac{m_{h}}{m_{X}} \right)^{2} } \;.
\end{eqnarray}
We show the branching ratios of the scalar boson $ X(750) $ 
with $ \Gamma_X = 45 GeV $ for the four most dominant modes 
$ YY $, $ W^{+}W^{-} $, $ ZZ $, and $ t\overline{t} $ in 
Fig.~\ref{X-BR}, where we have neglected the branching ratios of $B(X
 \to hh, hY)$ (we will come back to this point in the next section).

Since the scalar boson $ Y $ also mixes with the SM Higgs boson, it can 
decay into SM particles via the mixing. Thus, for the 
${\cal O}(1)$ GeV scalar 
boson $ Y $, the dominant decay modes are $Y \to \ell^+ \ell^-$ $(\ell=e,\mu)$
and $Y \to \pi \pi$ given by \cite{Gunion:1989we}
\begin{eqnarray}
\Gamma (Y \to \ell^+ \ell^-) &=& \sin^2\theta_{2} \, 
  \frac{m_\ell^2 m_{Y}}{ 8 \pi \langle \phi \rangle^2 } \left( 1 - 
  \frac{4 m_\ell^2}{m_{Y}^2 } \right )^{3/2}\;, \label{15}\\
\Gamma (Y \to \pi \pi) &=& \sin^2\theta_{2} \, 
  \frac{m_{Y}^3}{ 216 \pi \langle \phi \rangle^2 } \,
 \left( 1 -   \frac{4 m_\pi^2}{m_{Y}^2 } \right )^{1/2} \,
 \left( 1 +  \frac{11 m_\pi^2}{2 m_{Y}^2 } \right )^2 \;, \label{16} \\
\Gamma_{Y} &=& \frac{1}{\tau_{Y}} = \sum_{\ell = e, \mu} 
\Gamma(Y \to \ell^+ \ell^- )
          +  \sum_{\pi= \pi^+,\pi^0} \Gamma(Y \to \pi \pi )  \;, \label{17}
\end{eqnarray}  
Here $\pi\pi$ includes $\pi^+\pi^-$, $\pi^0\pi^0$ and $\Gamma
(Y\to\pi^{+}\pi^{-})=2\Gamma (Y\to\pi^{0}\pi^{0})$. Since the
tree-level estimate of $\Gamma (Y \to \pi \pi)$ is not correct 
when $m_Y$ is not far from the pion threshold, 
where strong final-state interaction becomes important 
\cite{Donoghue:1990xh,Bezrukov:2009yw}, so we
follow Ref.\cite{Donoghue:1990xh,Bezrukov:2009yw} for
numerical estimates of $\Gamma (Y \to \pi \pi)$. For a 1 GeV scalar
boson $ Y $, the branching ratio into $\pi\pi$ is almost $ 100\% $ and
for $\mu^+ \mu^-$ is just about $ 0.4\% $ \cite{Bezrukov:2009yw}.
The lifetime of $Y$ is equal to the inverse of the total width of
  $Y$. The total width is calculated by summing all the partial widths
  given in Eqs.~(\ref{15}) and (\ref{16}). For a 1 GeV scalar boson $Y$ with 
 $\sin\theta_{2}=1.6\times 10^{-2}$, we have $ \Gamma_{Y}\approx
  4.25\times 10^{-10} $ GeV and so $ \tau_{Y}=\frac{1}{\Gamma_{Y}}\approx
  1.55\times 10^{-15} $ (s).

The production cross section of $X(750)$ via gluon fusion is simply 
given the gluon fusion cross section of a would-be 750 SM Higgs boson
multiplied by the factor $\sin^2\theta_1$ as 
 \begin{equation}
  \sigma (pp \to gg \to X(750) ) = \sin^2 \theta_1 \times 
  \sigma_{\rm SM} (pp\to gg \to H_{\rm SM} ) \;.
\end{equation}

\section{Constraints}

\subsection{Constraints on $X$}
The scalar boson $X(750)$ mixes with the SM Higgs boson, and so it will
affect the Higgs boson data collected at the Run-1. Indeed, the $X(750)$
interacts with the SM particles via the mixing with the SM Higgs boson
with the angle $\theta_1$.
In this setup, being similar to the Higgs-portal models, a previous 
global fit to all the Higgs-boson data was performed \cite{Cheung:2015dta}.
The mixing angle $\theta_1$ is constrained to be, at 95\% CL 
\cite{Cheung:2015dta}.
\begin{equation}
\label{86}
 \cos\theta_1 > 0.86\,, 
\end{equation}
which implies $ | \sin\theta_1 | < 0.51$ or $\sin^2 \theta_1 < 0.26$.
In the following, we shall use a moderate value for 
$ | \sin\theta_1 |  = 0.3$ unless stated otherwise.

There are also other direct searches for new resonances decaying into 
$W^{+}W^{-}$, $ZZ$, $t\bar{t}$, $hh$, and dijets with the 8 TeV
data by both the ATLAS and CMS collaborations.
\begin{itemize}
\item The searches for a scalar resonance decaying to $ ZZ $ and $ WW$ 
 with the full data set exist for both ATLAS \cite{Aad:2015kna,Aad:2015agg} and 
 CMS \cite{Khachatryan:2015cwa}. Combining all relevant decay modes of 
  $Z$ and $W$, the $ 95\% $ CL upper limit on the production cross sections 
 are 
\begin{eqnarray}
\sigma(pp\rightarrow S)_{\rm 8 TeV}\times B(S\rightarrow ZZ) < 22 \;
  {\rm fb}_{\rm (ATLAS)},\;\; 27 \; {\rm fb}_{\rm (CMS)} \\
\sigma (pp\rightarrow S)_{\rm 8TeV}\times B(S\rightarrow W^{+} W^{-} ) < 
  38\; {\rm fb}_{\rm (ATLAS)},\;\;  220 \; {\rm fb}_{\rm (CMS)}.
\end{eqnarray}
For $ \sin \theta_{1}=0.3 $, with $\sigma_{\rm SM} (pp\to gg \to H_{\rm SM} )
\approx 157$ fb at $\sqrt{s}=8$ TeV and the branching ratio of
$X(750)$ into $WW$, $ZZ$, and $t\bar t$ can be found in Table~\ref{X-decay}, 
\begin{eqnarray}
\sigma(pp\rightarrow X)_{\rm 8TeV}\times 
  B(X\rightarrow ZZ) &=&  2.0 \; {\rm fb}
    \\
\sigma(pp\rightarrow X)_{\rm 8TeV}\times B
     (X\rightarrow W^{+}W^{-}) &=& 4.1  \;
  {\rm fb}
\end{eqnarray}

\item The searches for a scalar resonance decaying to a pair of 
 top quarks for both ATLAS \cite{Aad:2015fna} and 
  CMS \cite{CMS-tt} impose a $ 95\% $ CL upper limit of 
\begin{eqnarray}
\sigma(pp\rightarrow S)_{\rm 8TeV}\times B( S\rightarrow t\bar{t}) < 0.7 \;
{\rm pb}_{\rm (ATLAS)},\;\; 0.6\; {\rm pb}_{\rm (CMS)}.
\end{eqnarray}
For $\sin \theta_{1}=0.3 $, $ \sigma(pp\rightarrow X)_{\rm 8TeV}\times 
 B(X\rightarrow t\bar{t}) = 0.38 fb $.
\item 
The searches for a scalar resonance decaying to a Higgs boson pair 
 for both ATLAS \cite{Aad:2015xja}  
 and CMS \cite{Khachatryan:2015yea} impose a $ 95\% $ CL upper limit of 
\begin{eqnarray}
\sigma(pp\rightarrow S)_{\rm 8TeV}\times B(S\rightarrow hh) < 35 \; 
 {\rm fb}_{\rm (ATLAS)},\;\; 52\; {\rm fb}_{\rm (CMS)}.
\end{eqnarray}
By imposing this constraint, we have
\begin{eqnarray}
|\mu_{Xhh}|\lesssim 435 \; {\rm GeV} \;.
\end{eqnarray}
Because $ |\mu_{Xhh}| $ is a free parameter in our model and not 
relevant to explain the di-photon resonance, 
let us assume $ Br(X\rightarrow hh)\ll\mathcal O(1\%)  $ {(i.e. $ |\mu_{Xhh}|\lesssim 190 GeV $)} and ignore its effect below.

Since the leading term in ${\cal L}_{Xhh}$ in Eq.~(\ref{LXhh}) and
${\cal L}_{XYY}$ in Eq.~(\ref{LHD}) are not suppressed by any of the
mixing angles $\sin\theta_i \, (i=1,2,3)$ while the leading term for
${\cal L}_{XYh}$ derived from the Lagrangian contains at least one
factor of $\sin\theta_i$, and therefore the $XYh$ coupling will be
suppressed relative to $Xhh$ and $XYY$. In the computation of
branching ratios of $X$, we shall also ignore $X \to hY$, as we have
assumed $B(X \to hh) \ll {\cal O}(1\%)$.

\item 
The searches for a scalar resonance decaying to dijet also appear for
both ATLAS \cite{Aad:2014aqa} and CMS \cite{CMS-dijet}. Here since we
produce $X(750)$ by gluon fusion, we also consider the $ 95\% $ CL
upper limit on the production of a RS graviton decaying to gg from CMS
\cite{CMS-dijet}:
\begin{eqnarray}
\sigma(pp\rightarrow S)_{\rm 8TeV}\times B( S\rightarrow gg)\times\alpha < 1.8 \;
{\rm pb}.
\end{eqnarray}
where $\alpha$ is the acceptance.  For $\sin \theta_{1}=0.3 $ and
conservatively setting $\alpha$=1, $ \sigma(pp\rightarrow X)_{\rm
  8TeV}\times B(X\rightarrow gg) = 1.76\times 10^{-3} fb $. 
\end{itemize}

\subsection{Constraints on $Y$}
So far, we have discussed the constraints related to the heavy resonance
of 750 GeV. The very light boson $Y$, which has mass of order 1 GeV, 
is also subject to a number of constraints as follows.

\begin{itemize}
\item
The boson $Y$ mixes with the SM Higgs boson via the mixing angle 
$\theta_2$, which is very similar to the mixing between $X$ and the
SM Higgs. Thus, the constraint on $\theta_2$ from the Higgs boson data is
the same as $\theta_1$:
\[
   \cos \theta_2 > 0.86 \qquad \mbox{the same as Eq.~(\ref{86})} \;.
\]
Since $h$ also decays into $YY$ via the mixing angle $\theta_2$, 
the size of $\theta_2$ is then carefully chosen to be
consistent with the Higgs boson data in the same way as $\theta_1$.

\item
Another set of constraints on $\sin\theta_2$ arise from $B$ and $K$ 
meson decays \cite{clarke}
For $360\, {\rm MeV} < m_Y < 4.8 \,{\rm GeV}$, 
the strongest constraint comes from
$B \to K \mu^+ \mu^-$ decay limit, which constrains
$\sin^2 \theta_2 \times  B(Y \to \mu^+ \mu^-) < 10^{-6}$, which implies 
$\sin^2 \theta_2 \times 0.4 \% < 10^{-6}$ for $m_Y = 1$~GeV.
Therefore, $\sin\theta_2 < 1.6 \times 10^{-2}$. 

\item
The $\theta_2$ is also constrained by more specific searches of the
Higgs boson decays, such as $h \to a a \to (\gamma\gamma)(\gamma\gamma)$. The
closest one that we can find is by the ATLAS Collaboration \cite{3gamma}.
In the paper, one of the searches is $h \to a a \to (\gamma\gamma)
(\gamma \gamma)$, but only for $10 \, {\rm GeV} < m_a < 62\,{\rm GeV}$. 
In this case,
the photon pair in each decay of the boson $a$ is widely separated
and can be detected.  It is very different from our case of $Y$ being
around 1 GeV such that the photon pair is really collimated.

\item

There was another search by ATLAS \cite{aaa} in
\[
    h (125) \to a a \to (\gamma \gamma) (\gamma \gamma) \;,
\]
for $m_a = 100, 200, 400$ MeV. In this case, the photon pair from each
boson $a$ is very collimated and could not be distinguished from
a single photon. ATLAS then used $h \to \gamma\gamma$ to constrain
this rare decay and set a limit:
\[
   B(h \to a a ) \times B^2\, (a \to \gamma \gamma) < 0.01 \;.
\]
However, the range of the boson $Y$ that we are using in this work is
$O(1\,{\rm  GeV})$ such that most of the range is not covered by this
current limit.  Thus, it is important to set
up specific high-angular-resolution search for the case of $m_Y \sim 
{\cal O}(1 \, {\rm GeV})$.
\end{itemize}

For the mass range of $Y$ from 0.5 GeV to 1 GeV, the major decay mode is 
$\pi \pi$, which is larger than 90\%, and the sub-leading decay mode is 
$\mu^+ \mu^-$, which is less than 10\%. We choose $m_Y=1$~GeV as our 
benchmark point because it can give $B(Y \to \pi \pi) \sim 100\%$ and 
$B(Y \to \mu^+ \mu^-) \sim 0.4\%$ . On the other hand, 
there is no reliable calculation for the hadronic branching ratios for
$m_Y$ between 1 GeV to 2.5 GeV. 
Finally, for $m_Y > 2.5$ GeV the decay mode of $\pi \pi$ becomes negligible.
Therefore, we focus on the mass of $Y$ to be $\mathcal{O}(1\, {\rm GeV})$.

In summary, for $m_Y \sim O(1\,{\rm GeV})$ the angle $\theta_2$ is constrained
to be $\sin\theta_2 < 1.6 \times  10^{-2}$. 
We shall then use $\sin\theta_2 = 1.6 \times  10^{-2}$ 
and $m_Y = 1$ GeV for numerical results.
With this choice of $\sin\theta_2$
the $h \to YY$ easily satisfies all the constraints.

\section{Fitting the Di-photon Resonance}

\begin{figure}[t!]
\centering
\includegraphics[width=3.5in]{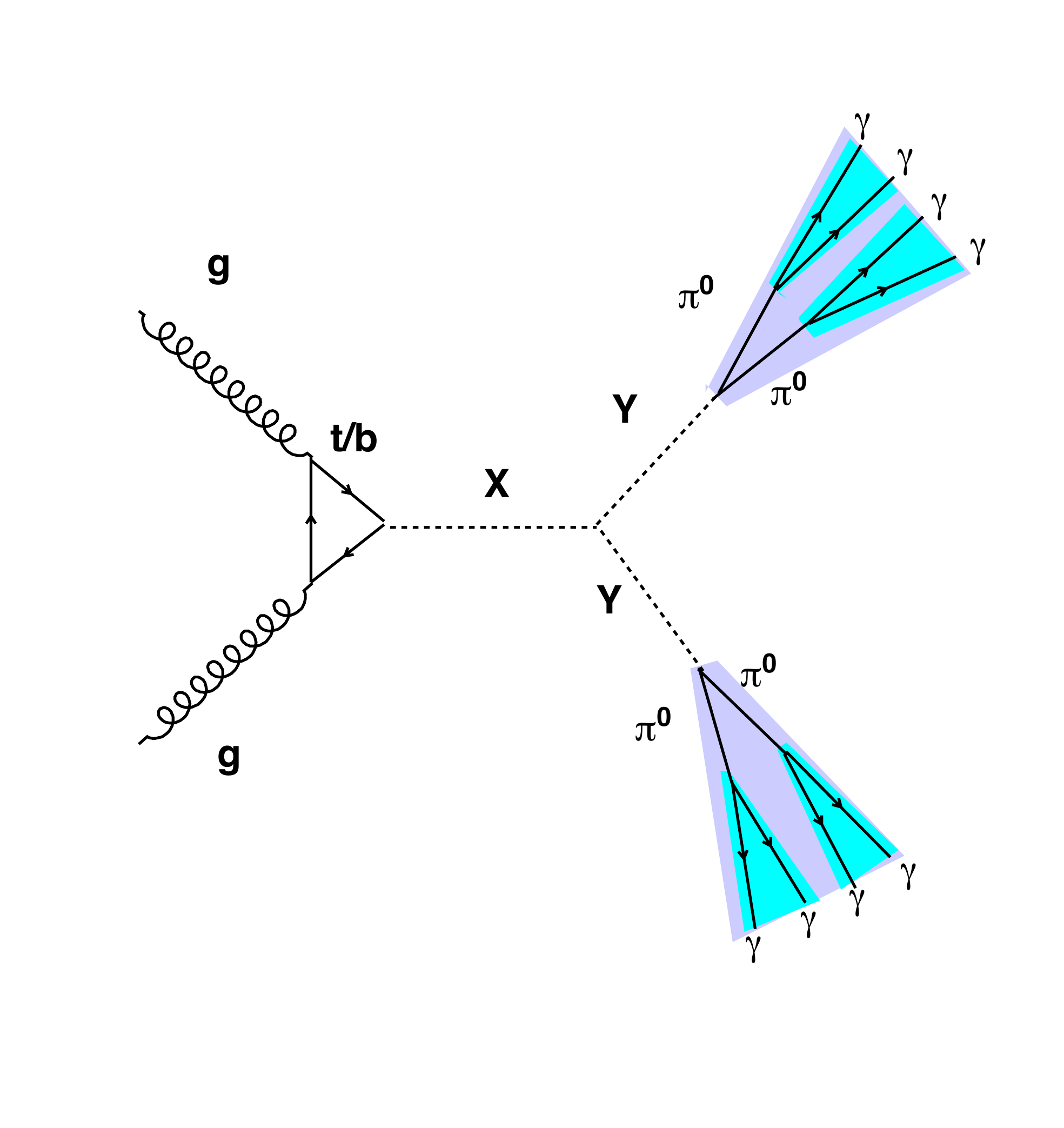}
\caption{\small \label{X-8a}
The Feynman diagram for $ pp\rightarrow X\rightarrow YY\rightarrow 
(\pi^{0}\pi^{0})(\pi^{0}\pi^{0})\rightarrow (4\gamma)(4\gamma) 
$ (2 $\gamma$-jets in the final state)}
\end{figure}

To fit the di-photon resonance, we consider the following process
\begin{equation}
pp\rightarrow X\rightarrow YY\rightarrow (\pi^{0}\pi^{0})(\pi^{0}\pi^{0}) 
 \to (4\gamma) ( 4\gamma) \;.
\end{equation}
The Feynman diagram for this process is shown in Fig.~\ref{X-8a}.
Since each $ \pi^{0} $ almost $100\%$ decays to $ \gamma\gamma $, the
final state consists of a total 8 photons.  Because these photons come from the
decay of a fast moving light particle $ Y $, they form a cluster of
collinear photons (which we called photon-jet ($ \gamma $-jet)) on
each side. Because the photons inside each photon-jet are so
collimated that they cannot be distinguished from a single photon, we 
use this feature to interpret the di-photon excess. Before discussing
how the photon-jet can escape from the experimental isolation conditions,
let us first fit the observed di-photon resonance width and {production rate}.
\subsection{Fitting the width for $X(750)$}

\begin{figure}[t!]
\centering
\includegraphics[width=3.5in]{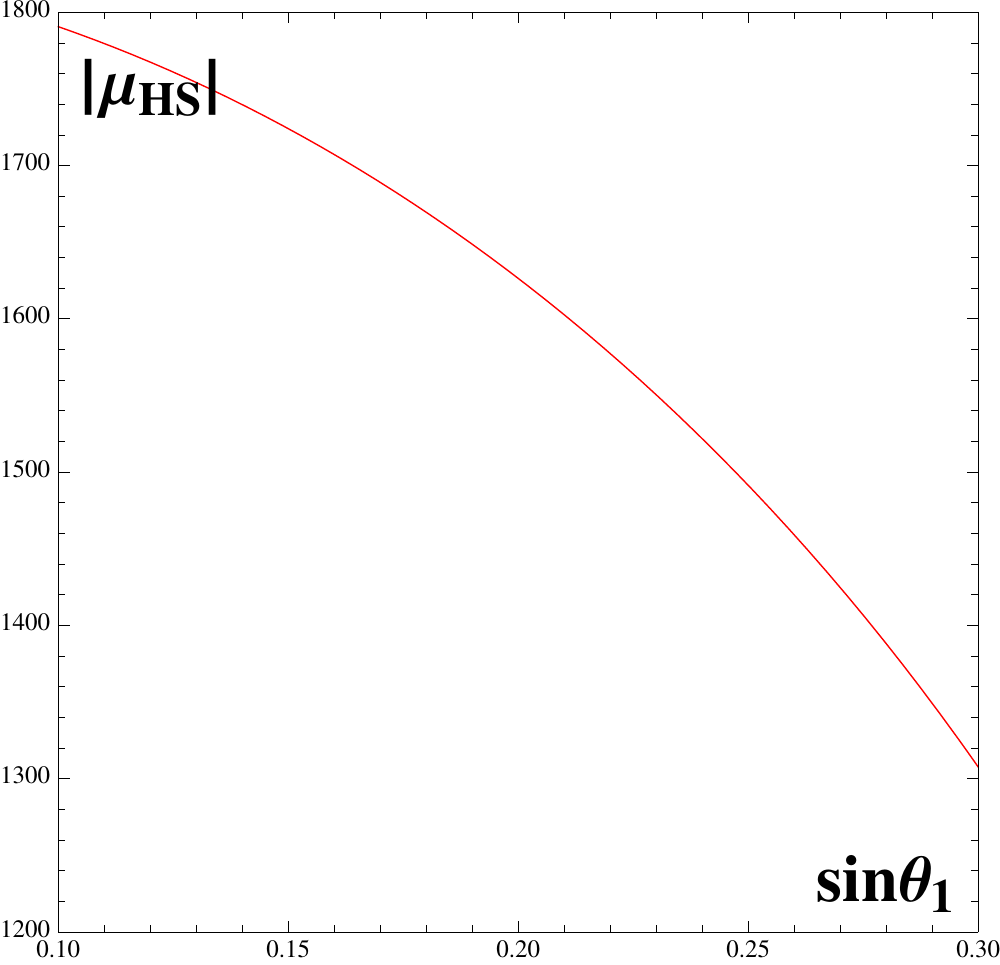}
\caption{\small \label{param-1}
The contour of $\sin\theta_{1} $ versus $ |\mu_{HS}| $ 
for a fixed $ \Gamma_{X(750)} = 45$ GeV and $ m_{Y}= 1$ GeV. }
\end{figure}

In order to fit the width of $X(750)$ to 45 GeV with the mixing angle 
$ \sin\theta_{1}=0.3 $, we need a very strong coupling 
for $ X $ with a pair of $ Y $'s
\begin{eqnarray}
|\mu_{HS}|\gtrsim 1308 \; {\rm GeV} \;.
\end{eqnarray}
We have pointed out in the Introduction that because the width
relative to the mass of the resonance is rather large ($\Gamma/M\approx 6\% $),
it implies that the new particle must couple
strongly to its decay products. 
The branching ratios and partial widths for $X(750)$ into 
the four most dominant modes $ YY $, $W^{+}W^{-} $, $ ZZ $, and 
$t\bar{t} $ are shown in Table~\ref{X-decay}. 
We also show in Fig.~\ref{param-1} the contour of $ \sin\theta_{1}$ 
versus $|\mu_{HS}|$ where we fix $\Gamma_{X(750)} = 45$ GeV.
The contour indicates that the weaker the mixing angle $ \sin\theta_{1} $, 
the stronger coupling of $ X\rightarrow YY $ required.
The lower region of the contour corresponds the total width $\Gamma_{X(750)}
< 45$ GeV.

\begin{table}[h!]\small
 \centering
  \begin{tabular}{ccccc}
\hline
& $YY$ & $W^{+}W^{-}$ & $ZZ$ & $t\overline{t}$ \\
\hline
BR & $50.45\%$ & $29.07\%$ & $14.38\%$ & $6.10\%$ \\
$ \Gamma_{i} $ (GeV) & 22.70 & 13.08 & 6.47 & 2.75 \\ \hline
   \end{tabular}
   \caption{\small \label{X-decay}
The branching ratios and partial widths for $X(750)$ into
the four most dominant modes $ YY $, $W^{+}W^{-} $, $ ZZ $, and $t\bar{t} $ 
with the mixing angle fixed at $\sin\theta_{1}=0.3 $.
}
\end{table}

\subsection{Fitting the production rate for $X(750)$}

\begin{figure}[t!]
\centering
\includegraphics[width=3.5in]{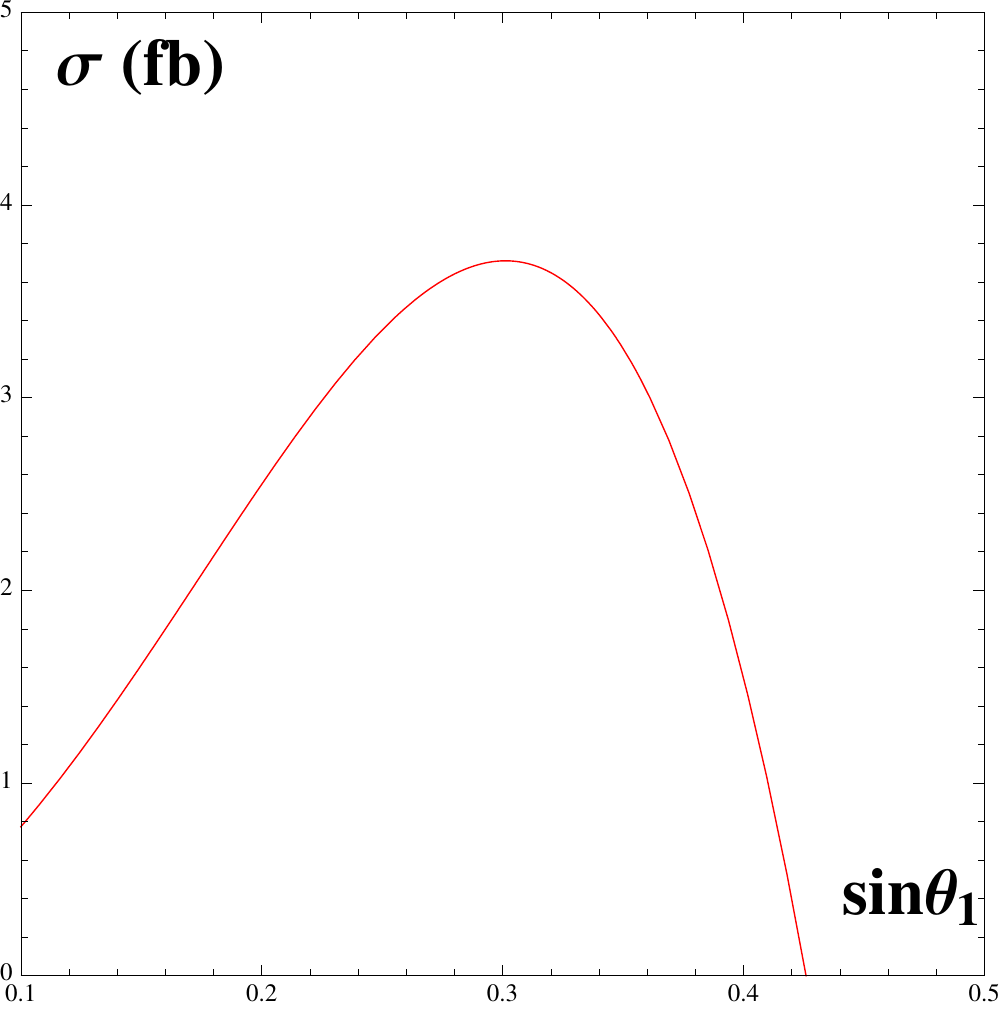}
\caption{\small \label{Xsec}
The variation of production rate of
$\sigma \left (pp\rightarrow X\rightarrow YY\rightarrow 
 (\pi^{0}\pi^{0})(\pi^{0}\pi^{0})\rightarrow (4\gamma)(4\gamma) \right )$
 versus  $\sin\theta_{1}$}
\end{figure}

To fit the production rate, we consider the cross section for 
$pp\rightarrow X\rightarrow YY\rightarrow (\pi^{0}\pi^{0})(\pi^{0}\pi^{0})
 \rightarrow (4\gamma)(4\gamma) $
\begin{eqnarray}
&&\sigma \left (pp\rightarrow X\rightarrow YY\rightarrow 
 (\pi^{0}\pi^{0})(\pi^{0}\pi^{0})\rightarrow (4\gamma)(4\gamma) \right )
\nonumber \\
&& =\sigma (pp\rightarrow X) \times B(X\rightarrow YY)
 \times \left[ B(Y\rightarrow \pi^{0}\pi^{0}) \right ]^{2} 
 \times \left[ B(\pi^{0}\rightarrow\gamma\gamma)\right ]^{4}
\nonumber \\
&&\approx \left[736\; {\rm fb} \times (0.3)^{2} \right] 
  \times [50.45\%] \times \left[100\%\times\frac{1}{3} \right]^{2} 
   \times [100\%]^{4} \nonumber \\
&&\approx 3.71 fb 
\end{eqnarray}
where the gluon fusion production cross section at $ \sqrt{s} = 13$ TeV
for a SM Higgs boson of mass $ M_{H} = 750$ GeV  is 
$\sigma(gg\rightarrow H_{SM})\approx 0.736$ pb \cite{ggF}.
We also show the variation of production rate versus
 $\sin\theta_{1}=0.1 - 0.5$ in Fig.~\ref{Xsec}. 
We can see the maximum cross section that we can obtain is about 4 fb and 
it occurs at around $\sin\theta_{1}\approx 0.3$.

\subsection{Discussion of the photon-jet scenario}

Since the photon-jet can arise from a highly boosted object that
decays 
into multiple photons, which merge into a
photon-jet appearing like a single photon,
which then hit the electromagnetic
calorimeter (ECAL) at essentially the same place. In our Hidden-Valley-like 
simplified model, the $X(750)$ decays into a pair of $Y$. 
Each $Y$ decays into two neutral pions and each neutral 
pion further decays into 2 photons.
Therefore, the final state consists of 
a back-to-back pair of photon-jets, each of which consists of 
4 collimated photons.
Due to very tiny angular separation among the photons in each photon-jet
and the long lifetime, the $\gamma$-jet may be 
reconstructed as a single photon.

The ATLAS and CMS have different capabilities in both detecting
photons and distinguishing them from other possible electromagnetic
objects. Let us briefly summarize them here
\cite{atlas-photon,cms-photon}. The granularity of CMS ECAL is
$0.0174\times 0.0174$ in $\eta \times\phi$.  The ATLAS electromagnetic
calorimeter consists three longitudinal layers. The granularity for
the first layer (a thickness between 3 and 5 radiation lengths) is
$0.003-0.006$ in $\eta$ within the regions $1.4 < |\eta| < 1.5$ and
$|\eta| > 2.4$, sufficient for the photons decaying from $\pi^0$. The
second layer (thickness around 17 radiation lengths) granularity is
$0.025\times0.025$ in $\eta\times\phi$.  The third layer is for high
energy shower correction.

Let us perform a simple estimate for 
the angular separation and the lifetime of $Y$
in order to confirm this scenario. 
The angular separation in the decay products of $Y$ is roughly
$\triangle R\approx 2m_{Y}/P_{T_{Y}}=2 {\rm GeV}/375 {\rm GeV} = 0.0053$. The
lab-frame lifetime of $Y$ is
$\gamma\tau_{Y} \approx 2 \tau_{Y}/\triangle R \approx 5.85\times 
10^{-13}(s)$, so the decay length could be $\gamma
c\tau_{Y}\approx 0.18 ({\rm mm})$, 
where $\gamma$
and $ c $ are the boost factor and the speed of light,
respectively \cite{Agrawal:2015dbf}.
As we know, if the decay length of a particle is less than 
0.15 mm, then we can take it as a prompt decay. Even though 
the boson $Y$ in our scenario is a rather long-lived particle, 
its decay length is very close to the prompt decay.

The high-granularity calorimeter of ATLAS is capable of resolving
single photons from $\pi^0$s. However, they used an elliptical cone
$(\Delta\eta/0.025)^2 + (\Delta\phi/0.05)^2 < 1$ (corresponding
roughly to a $3 \times 5$ cluster) to associate the photon
candidate
\footnote{\label{select-efficiency} For the photon selection
  efficiency, the ATLAS experiment selects the photon candidate with
  an elliptical cone, inside $(\Delta\eta/0.025)^2 +
  (\Delta\phi/0.05)^2 < 1$ (corresponding roughly to a $3 \times 5$
  cluster) the true photon with highest $p_T$ is associated to the
  candidate \cite{atlas-photon}. They extrapolated the ``final state''
  particle to their impact point in the second sampling of the
  electromagnetic calorimeter (EMS2). The reconstructed photon and the
  true particle association are depending on the difference between the
  position $(\eta^{clus},\phi^{clus})$ of the cluster barycentre and
  the coordinates $(\eta^{extr}, \phi^{extr})$, $\Delta \eta =
  \eta^{extr} -\eta^{clus}$ and $\Delta \phi = \phi^{extr}
  -\phi^{clus}$.}, 
which might take the highly boosted photon-jet as a single photon. 

The LHC experiments in the Higgs measurements of the di-photon
channel are able to distinguish photons of $m_h/2 \sim 65$ GeV from
$\pi^0$'s of the same energy \cite{Knapen:2015dap}.  
The angular separation between the
two photons in the 65 GeV pion decay is roughly $0.004$.  The efficiency
also depends strongly on the direction of the pions.  Since $X \to YY
\to (\pi^0 \pi^0) (\pi^0 \pi^0)$, so that each $\pi^0$ has a maximum
energy of $187.5$ GeV, which corresponds to an angular separation
between the pair of photons from $\pi^0$ of about $0.0015$, which is
about a factor of 3 smaller. When we go one step back in the decay
chain, the angular separation between two neutral pions from each $Y$
decay is about $2 m_Y / P_T \simeq 0.005$.  The overall picture is
exactly what we showed in Fig.\ref{X-8a}. Each $Y$ decays into $2 \pi^0$ with
an angle $0.005$, and each $\pi^0$ decays into 2 photons with an angle
$0.0015$. All 4 photons are contained in a cone of 0.005 forming a
photon-jet. Certainly
without a dedicated high-photon-resolution analysis, it is very
difficult to distinguish a photon-jet from a single photon in this
case. In the limited information given in the proceedings
\cite{ATLAS-750,CMS-750}, we cannot tell if the signal contains single
photons or photon-jets.  A more elaborate experimental analysis is
needed to determine if it is possible.

Finally, our scenario is still valuable to make some
predictions even without a detailed detector simulation. Let us
discuss it in the following 3 points.

\begin{itemize}
\item First, since the angular separation between the pair of photons
  from $\pi^0$ in our scenario is just about a factor of 3 smaller
  than that between the photon pair from the decay of a neutral pion
  of energy 65 GeV, it could be resolved for either ATLAS
  or CMS in the near future, similar to the case between the 65 GeV photon
  and a neutral pion of same energy. We have shown a clear
  physical picture in Fig.\ref{X-8a}.

\item Second, it will be much more plausible to detect the photon-jet
  from the prompt decay of $Y$ rather than the non-prompt case.The
  decay length of $Y$ in our scenario is only 0.18 mm
  which is not far from the prompt decay. We suggest that
    both ATLAS and CMS can have a higher chance to distinguish the
    photon-jet from the single photon in our scenario than the other
    proposals, e.g. in Ref.~\cite{Agrawal:2015dbf}, of longer decay
    lengths of very light particles with seriously non-prompt decays.
  These kind of seriously non-prompt decays will become virtually
  impossible to make any quantitative predictions without detailed
  detector simulations, but our scenario is almost free from the
  complication of the non-prompt decays.

\item Third, if the experimental groups have fine enough resolution to
  distinguish the photon-jet from the single photon in our scenario,
  then they can also clearly observe a resonance around 1 GeV.
\end{itemize}
Based on these three points, our scenario can explain why
the photon-jet is still indistinguishable from the single photon under
the current experimental studies and the current constraints.
However, it is still plausible to explore the difference between 
photon-jets and single photons in our scenario with $m_Y \approx 1 $ GeV.
Therefore, we strongly encourage our experimental colleagues to
perform a more elaborate experimental analysis to confirm or
rule out this kind of scenario.

\section{Discussion}
We have studied the 750 GeV di-photon resonance by interpreting it as 
the production of 
the resonance $X(750)$ decaying into a pair of very light 
($\mathcal{O}(1\,{\rm  GeV})$) 
particles $Y$, each
of which in turn decays to 4 photons which form photon-jets. Since these
photons inside the photon-jet are so collimated that they cannot be
distinguished from a single photon. 
So far, we have found that this
scenario can simultaneously accommodate the width and the production rate
though moderately, and is consistent with 
the current constraints from the 8 TeV searches and the isolation of 
photon-jets from single photons.

We offer the following comments with regards to our scenario. 
\begin{enumerate}

\item 
There are also two other processes with the production rate of order 
$\mathcal{O}(15\;{\rm fb})$ in our scenario:
\begin{equation}
pp\rightarrow X\rightarrow YY\rightarrow (\pi^{0}\pi^{0})(\pi^{+}\pi^{-}) 
\label{npcp}
\end{equation}
and 
\begin{equation}
pp\rightarrow X\rightarrow YY\rightarrow (\pi^{+}\pi^{-})(\pi^{+}\pi^{-}) \;.
\label{cpcp}
\end{equation}
The Feynman diagrams for these two processes are shown in 
Fig.~\ref{X-4ajj} and Fig.~\ref{X-jjjj}, respectively.
Since the two charged pions coming from $Y$ decay are very collimated, 
they will appear as a ``microjet'', which looks like 
a $\tau$-jet experimentally,  which is
rather ``thin'' compared to the usual hadronic jet \cite{Cheung:2013oya}.
The pixel detector inside the LHC experiments has some chances of
separating them \cite{pixel detector,Cheung:2013oya}.
If the experiment cannot resolve these two charged pions, then the final
state will consist of a ``microjet'' of two unresolved charged pions and
a photon-jet for Eq.~(\ref{npcp}) and two ``microjets'' 
for Eq.~(\ref{cpcp}).

\begin{figure}[t!]
\centering
\includegraphics[width=4in]{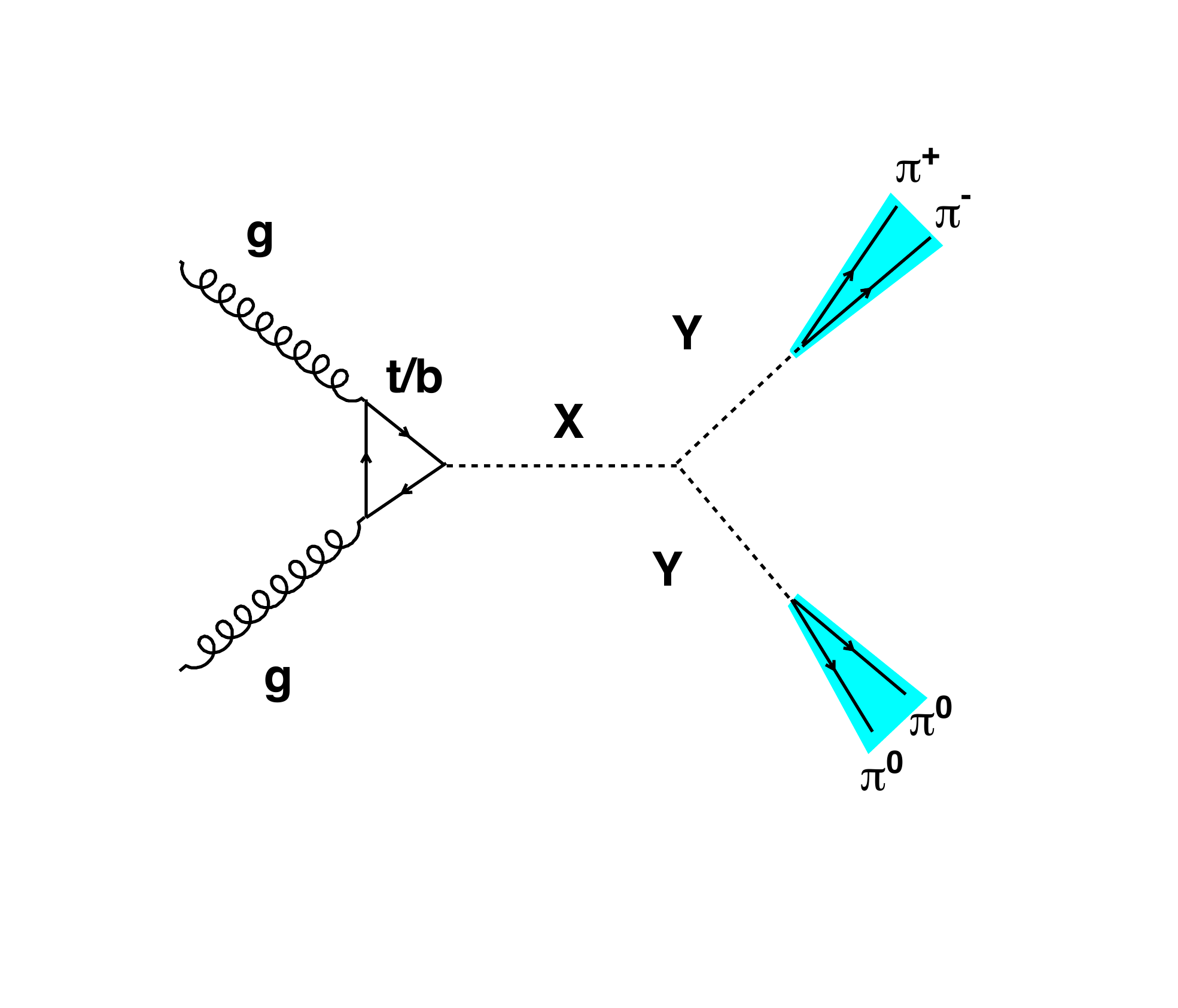} 
\caption{\small \label{X-4ajj}
The Feynman diagram for 
$ pp\rightarrow X\rightarrow YY\rightarrow (\pi^{0}\pi^{0})(\pi^{+}\pi^{-})$.}
\end{figure}
\begin{figure}[t!]
\centering
\includegraphics[width=4in]{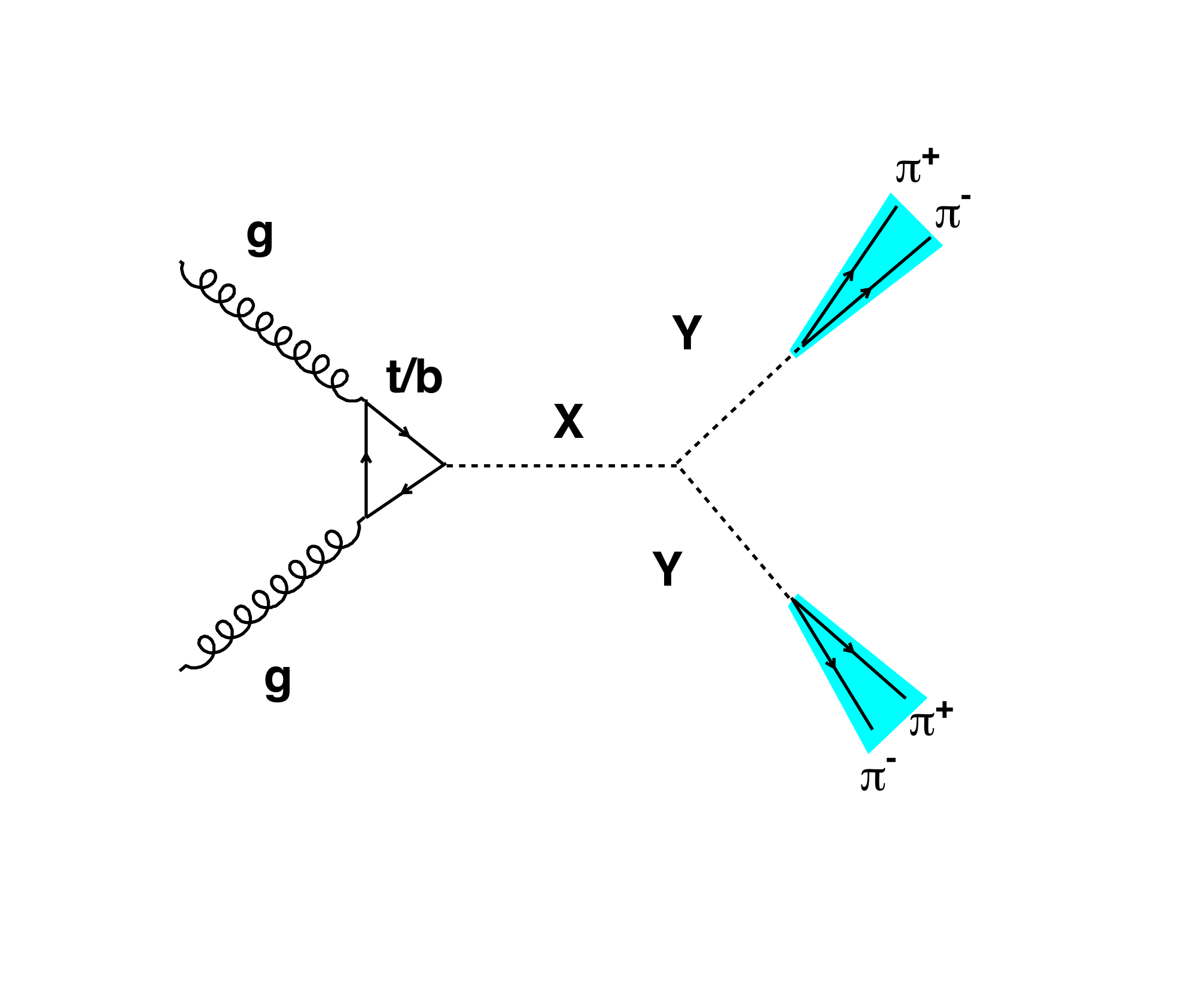} 
\caption{\small \label{X-jjjj}
The Feynman diagram for 
$ pp\rightarrow X\rightarrow YY\rightarrow (\pi^{+}\pi^{-})(\pi^{+}\pi^{-})$.}
\end{figure}

\item Another interesting process but with a much smaller production rate of 
order $\mathcal{O}(0.1 \,{\rm fb})$ in our scenario is
\begin{equation}
pp\rightarrow X\rightarrow YY\rightarrow (\pi^{0}\pi^{0})(\mu^{+}\mu^{-}) 
\label{npmu}
\end{equation}
The Feynman diagram for this process is shown in Fig.~\ref{X-4a2mu}.
The collimated muon-pair seems more possible to be resolved 
than the photon-jet or microjet \cite{muon-jet},
because of the much-refined pixel detector in the central region together
with the outer  muon chamber.
Thus, the final state will consist of a pair of collimated muons appearing
as a muon-jet and a photon-jet.  It is a rather clean signal to be identified. 

\begin{figure}[t!]
\centering
\includegraphics[width=3.5in]{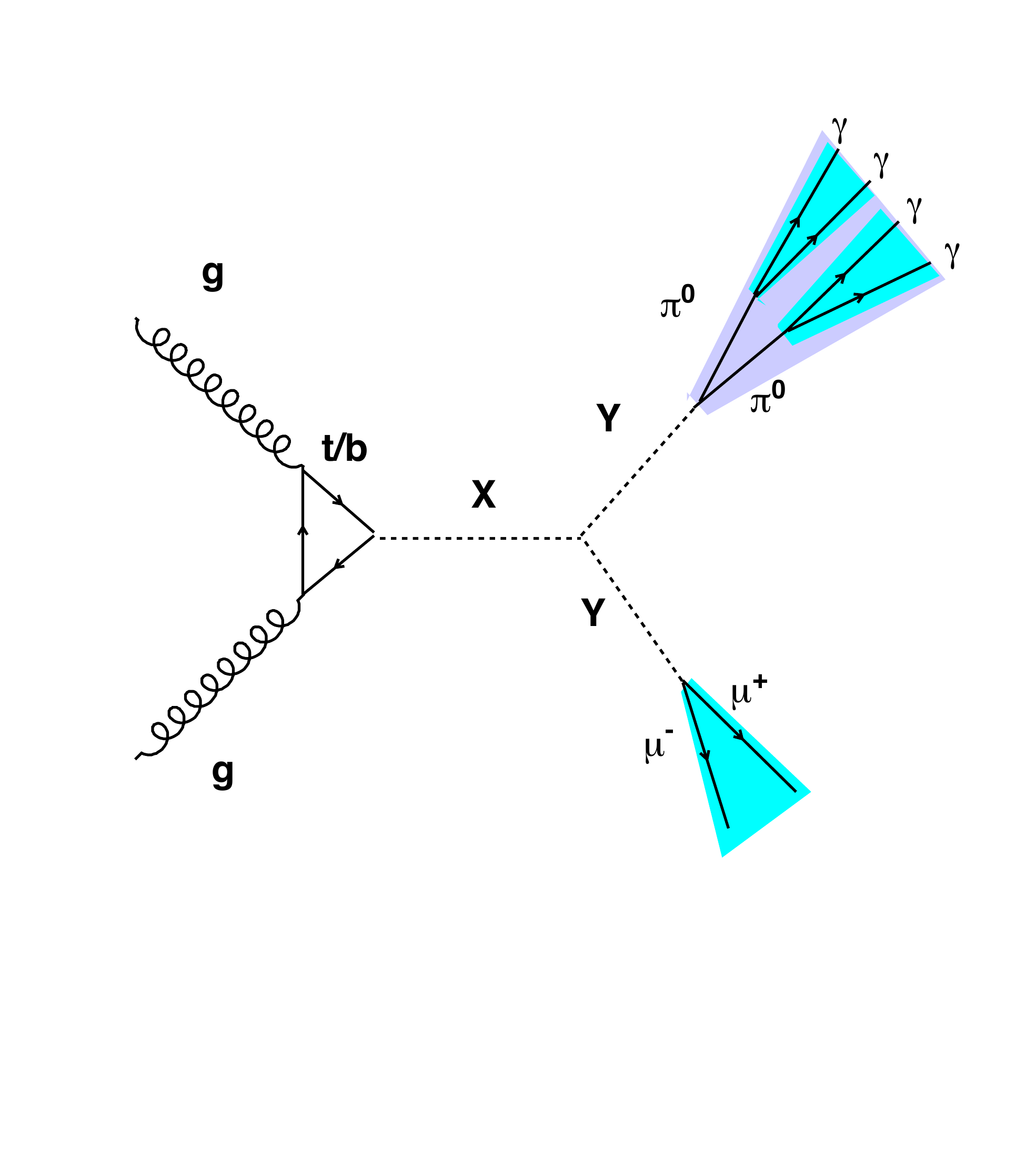}
\caption{\small \label{X-4a2mu}
The Feynman diagram for $ pp\rightarrow X\rightarrow YY\rightarrow 
(\pi^{0}\pi^{0})(\mu^{+}\mu^{-})\rightarrow (4\gamma)(\mu^{+}\mu^{-})$}
\end{figure}

\item 
Similar ideas using collimated photons to explain the di-photon resonance
also appeared in Ref.~\cite{Knapen:2015dap,Agrawal:2015dbf}. They
also used a scalar boson of 750 GeV, but the differences are 
as follows.
\begin{itemize}
\item In Ref.\cite{Knapen:2015dap}, they used a simplified model of
Hidden Valley. However, they considered a very light scalar 
with a new charged vector-like 
fermion at the weak scale to produce collimated pairs of photons.
They also focus on the
mass range of the very light scalar less than 1 GeV.

\item In Ref.\cite{Agrawal:2015dbf}, they used two types of models. The
first one is a very light CP-odd scalar axion (about 2 GeV) with a new
heavy lepton to produce collimated pairs of photons. The second one is
using ``fake'' photons by  adding new dark photons.
\end{itemize}
Comparing to these two works, our scenario is somewhat more economical
and offers different signatures. The new 750 GeV resonance is currently
detected by a pair of photon-jets, yet the scenario allows 
other interesting signatures to distinguish them.

\end{enumerate}

\section*{Acknowledgment}  
We sincerely thank the discussion with Jae-Sik Lee and Po-Yan Tseng.
The work of K.C. was supported by the MoST of Taiwan under Grants 
No. NSC 102-2112-M-007-015-MY3.

\end{document}